\documentclass[a4paper,10pt,twocolumn,showpacs,preprintnumbers,amsmath,amssymb]{revtex4}
\usepackage{graphicx}
\usepackage{dcolumn}
\usepackage{bm}
\usepackage{literat}
\usepackage{amsmath}

\begin{document}

\preprint{}
\title{Modeling self-organization of nano-size vacancy clusters\\ in stochastic systems subjected to irradiation}
\author{Dmitrii O.~Kharchenko}\email{dikh@ipfcentr.sumy.ua}
\author{Vasyl O.~Kharchenko}
\author{Anna I.~Bashtova}

 \affiliation{Institute of Applied Physics,
National Academy of Sciences of Ukraine, 58 Petropavlivska St., 40000 Sumy,
Ukraine}

\date{\today}
\begin{abstract} A study of the  self-organization of vacancy clusters in irradiated
materials is presented. Using a continuum stochastic model we take into account
dynamics of point defects and their sinks with elastic interactions of
vacancies. Dynamics of vacancy clusters formation is studied analytically and
numerically under conditions related to irradiation in both reactors and
accelerators. We have shown a difference in patterning dynamics and studied the
external noise influence related to fluctuation in a defect production rate.
Applying our approach to pure nickel irradiated under different conditions we
have shown that vacancy clusters having a linear size $\simeq 6~nm$ can arrange
in statistical periodic structure with nano-meter range. We have found that
linear size of vacancy clusters at accelerator conditions decreases down to
$20\%$, whereas a period of vacancy clusters reduces to $6.5\%$.
\end{abstract}

 \maketitle

\section{Introduction}
Metals and alloys under laser and particle irradiation are typical examples of
nonequilibrium systems manifesting self-organization processes of point defects
with nano-size patterns forming a corresponding microstructure. It is well
known that depending on irradiation conditions (displacement damage rate and
temperature) point defects can arrange into objects of higher dimension like
clusters (di-, tri-, tetra-vacancy clusters), defect walls with vacancy and
interstitial loops \cite{118,1110}, voids \cite{111}, precipitates \cite{117},
bubble lattices \cite{114,115,116}. Patterns with spatial structures of
nano-meter size can be observed as in a bulk as on a surface of irradiated
materials, for example, at ion-beam sputtering
\cite{CB95,MB2004,PRE2010,CMP2011,MetFiz2013}, condensation from gaseous phase
\cite{PhysScr2012,PRE2012} and molecular beam epitaxy
\cite{PhysScr2011,EPJB2013}. Theoretical and experimental investigations of
such dissipative structures are widely discussed in literature
\cite{Martin,116,1112,Ryazanov,AbromeitMartin1999,Sugakov95}. Numerical studies
of defect clusters formation allow one to set up conditions for different kind
of patterns emerging in diffusion processes and in processes of defects
interaction \cite{EPJB2012,CMP2013}. A production of defects and the resulting
microstructure can principally modify the physical and mechanical properties of
an irradiated material leading to swelling, hardening and embrittlement
\cite{BF2004,DKK2009,Hardening,Osetsky,Gavini}. Therefore, from practical and
theoretical viewpoints a study of microstructure of material, defects and their
arrangement into clusters is a quite urgent problem in modern material science
and statistical physics.

According to the standard theory of defects production (see
Refs.\cite{Walgraef93,Walgraef95,Walgraef,Walgraef03,mirz1}) a formation of
point defects is of thermo-fluctuation character; the probability of such
processes grows with temperature and irradiation flux. It can be realized also
at large defects density due to activation barrier height change for defect
formation related to elastic deformation of the medium caused by defects
presence. It is known that the thermo-fluctuation mechanism for defects
generation plays a central role in a structural disorder formation. In this
case a concentration of defects exceeds their equilibrium concentration by
several orders (concentration of nonequilibrium vacancies is $\simeq10^{17}\div
10^{20}~cm^{-3}$). These defects can migrate in solid, annihilate, collapse
into loops or interact with other defects forming spatial dissipative
structures. The last one is possible beyond some threshold defined by material
properties and irradiation conditions. In such a case a uniform distribution of
defects becomes unstable due to their self-organization. As far as a formation
of organized defect clusters requires the production of defects in cascades the
corresponding microstructure was observed under neutron or ion irradiation.
Experimentally (see, for example Ref.\cite{JES}) it was shown that under $3MeV$
protons irradiation at low doses ($\simeq 0.01~dpa$) a distribution of stacking
fault tetrahedrons in $Ni$ and $Cu$ is homogeneous, whereas at elevated doses
($\simeq0.1~dpa$) fluctuations of point defect clusters were observed. At doses
up to $0.65~dpa$ one gets well periodic structure of defect clusters. Moreover,
as was shown in Ref.\cite{Kiritani90} vacancy clusters in the form of stacking
fault tetrahedra can be formed from collision cascades at low temperatures when
vacancies do not make their motion by thermal activation. Formation of stacking
fault tetrahedron was experimentally observed even under electron irradiation
of foils \cite{Kiritani78,Kiritani00,SYMK2004}. It was shown that vacancy
clusters in electron irradiated metals, appear under the circumstances of local
enrichment of vacancies as the result of the behavior of interstitials atoms.
In experimental studies vacancy clusters have linear size on several
nano-meters (in the interval $2\div7~nm$ depending on irradiation conditions
and used material target) with $7\div 20$ vacancies inside clusters
\cite{SYMK2004}.

Understanding of the defect population dynamics and microstructure
transformations is thus of fundamental importance, not for its intrinsic
interest, but also for its technological significance. It allows one to
describe macroscopic ordering processes, phase separation in solids under
different conditions of external influence
\cite{Mazias,Kiritani,EnriqueBellon00,Enrique63,YeBellon04_1}. Despite the main
mechanisms of radiation damage in solids are known \cite{Kiritani91} nowadays
multiscale approaches treating the system under consideration at different
hierarchical levels (from atomic to mesoscopic level) are used to study
microstructure transformation in irradiated materials
\cite{MLN,NIMB2001,MMS04}. On a mesoscopic level such processes can be studied
within the framework of a rate theory where concentrations/densities of main
structural elements (defects, their clusters, sinks) are considered
\cite{Martin,AbromeitMartin1999,1112}. Usually such models take into account
defects production rate, their annihilation and diffusion of mobile species
(vacancies and interstitials) \cite{Walgraef95,Walgraef}. Unfortunately, a
generation of defects caused by elastic deformation of the medium and their
interactions caused by ``chemical'' forces are not considered as usual.
Moreover, as far as irradiation occurs at elevated temperatures (depending on
the irradiation conditions, namely, dose rate $K\sim
10^{-6}\div10^{-3}~dpa/sec$ and temperature $T\sim 500\div1000~K$) fluctuations
of defects production lead to stochastic dynamics of defect populations.
Properties of stochastic defects production in irradiated materials and
stochastic formation of stacking fault tetrahedrons (vacancy clusters) were
discussed in Refs.\cite{Kiritani,Kiritani_MCF97}. A stochastic production of
defects during irradiation influence was proposed in Ref.\cite{Yanovski}, where
it was assumed that the corresponding stochastic athermal mixing of atoms
caused by irradiation can be described as spatially correlated random process
giving the effective flux opposite to the diffusion one. This idea was
exploited studying ordering, patterning and phase separation in irradiated
materials \cite{EPJB2010,PhysA2010_1,CEJP2011,UJP2012}. In such a case the
corresponding fluctuations are caused by an external influence (energy
dispersion of particles in irradiation flux, dispersion in an incidence angle,
etc.) and are considered as an external noise. From the other hand in systems
with birth-and-death processes or ``chemical reactions'' (production of defects
and their annihilation) and diffusion there is another kind of fluctuation
source relevant to internal nature of the system. The corresponding internal
noise is defined by a heat bath. Its influence on arrangement of ensemble of
vacancies was studied theoretically in Refs.\cite{EPJB2012,CMP2013,UJP2013},
where it was shown that depending on internal noise properties different kinds
of vacancy structures emerge.

From both theoretical and practical viewpoints it is important that dynamics of
defects and, therefore, the corresponding microstructure are quite different at
conditions relevant to irradiation in reactors ($K=10^{-6}~dpa/sec$, $T=773~K$)
and in accelerators ($K=10^{-3}~dpa/sec$, $T=900~K$). Under reactor conditions
the principal role belongs to diffusion processes, whereas irradiation in
accelerators gives high damage rates with small contribution of diffusion
processes. It results in different spatial arrangement of defects in these two
cases. Therefore, the studying different physical processes of defects
arrangement in irradiated materials under different irradiation conditions is
actual. In this paper we are aimed to study dynamics of defects at their
self-organization under reactor and accelerator conditions considering behavior
of point defects and their sinks. Using the approach proposed in
Ref.\cite{Walgraef95} we take into account defects production caused by local
elastic deformation of material and stochastic effects induced by external
noise influence. In our study pure $Ni$ (as most wide-spread constructional
material in atomic energy) was selected as a reference system. We will show
that the external noise influence can be different in the case of irradiation
under reactor and accelerator conditions. In the first case it promotes large
difference in vacancy clusters occupation numbers comparing to the second case.
Moreover, it is able to reduce the linear size of vacancy clusters and a period
of their arrangement. It will be shown that both linear size of vacancy
clusters and their period are of several nano-meter range.

The paper is organized as follows. In Section II we present a generalized model
describing the main mechanisms of defects formation with two spatial scales
related to diffusion scale and defects interaction scale. In Section III we
consider a reduced one-component model analytically assuming constant sink
densities. Here we find critical values for both defect damage rate and
temperature limiting the patterning in the system. Section IV is devoted to
study a general model for pattern formation in system of defects numerically.
Here we discuss the difference in defects arrangement under reactor and
accelerator conditions. We conclude in Section V.

\section{Model}
In the framework of the rate theory evolution equations for populations of
vacancies $c_v$ and interstitials $c_i$ are of the form
\cite{Walgraef95,Walgraef}:
\begin{equation}\label{le1}
\begin{split}
 &\partial_t  c_{i}=K(1-\varepsilon_i)-D_{i}S_i c_{i}-\alpha c_{i}c_{v},\\
  &\partial_t c_{v}=K(1-\varepsilon_v)-D_{v}S_v(c_{v}-c_{0v})-\alpha
c_{i}c_{v}.
\end{split}
\end{equation}
Here the equilibrium vacancy concentration $c_{0v}=e^{-E^f_{v}/T}$ is defined
through vacancy formation energy $E^f_{v}$ and temperature $T$; $K$ is the
defect damage rate; $\varepsilon_{i}$ and $\varepsilon_{v}$ relate to cascades
collapse efficiency ($\varepsilon_v\gg \varepsilon_i$);
$D_{v,i}=D^0_{v,i}e^{-E^m_{v,i}/T}$ are diffusivities defined through the
migration energies $E^m_{v,i}$. Sink intensities $S_i\equiv
Z_{iN}\rho_N+Z_{iV}\rho_{v}+Z_{iI}\rho_i$ and $S_v\equiv
Z_{vN}\rho_N+Z_{vV}\rho_v+Z_{vI}\rho_i$ are determined by the network
dislocation density $\rho_N$ and densities of vacancy and interstitial loops
$\rho_{v,i}$ with preference $Z_{\{\cdot, \cdot\}\cdot}$ where
$Z_{vN}=Z_{vI}=Z_{vV}=1$, $Z_{iN}=1+B$, $Z_{iI}\simeq Z_{iV}\simeq1+B^{'}$; $B$
and $B^{'}\geq B$ are excesses for network bias and loop bias ($B\simeq 0.1$).
The recombination coefficient $\alpha={4\pi r_{0}(D_{i}+D_{v})}/{\Omega}$ is
defined through the interaction radius of defects $r_0$, the atomic volume
$\Omega$ and the corresponding diffusivities. Evolution equations for the loop
densities are as follows \cite{Walgraef95,Walgraef}:
\begin{equation}\label{le2}
\begin{split}
 &\partial_{t}\rho_i=\frac{2\pi N}{b}\left(\varepsilon_i K+D_iZ_{iI}c_i-D_vZ_{vI}(c_v-c_{v0})\right);\\
 &\partial_{t}\rho_v=\frac{1}{b r^0_v}\left(\varepsilon_v K-\rho_v[D_iZ_{iV}c_i-D_vZ_{vV}(c_v-c_{v0})]\right).
\end{split}
\end{equation}
Here $\vec b$ is the Burgers vector with modulus $b\equiv|\vec b|$, $r^0_v$ is
the initial vacancy loop radius, $N$ is the sinks density.

Using renormalized quantities $\rho'_{v,i}\equiv\rho_{v,i}/\rho_{N}$,
$t^{'}\equiv t \lambda_{v}$, $\lambda_{v}\equiv D_{v}Z_{v N}\rho_{N}$,
$x_{i,v}=\gamma c_{i,v}$, $\gamma\equiv \alpha/\lambda_{v}$, $P\equiv \gamma
K/\lambda_{v}$, $\mu\equiv(1+\rho'_{v}+\rho'_{i})$, $Z_{i N}/Z_{v N}=1+B$ and
introducing small parameter $\epsilon\equiv D_{v}/D_{i} \ll 1$ one can
eliminate population of interstitials adiabatically taking
$\epsilon\partial_{t}x_{i}\simeq 0$. It gives
$x_{i}={{P}(1-\varepsilon_i)}/[(1+B)\mu/{\epsilon}+ x_{v}]$. In such a case we
arrive at the system of three equations: one for the vacancy population
$x\equiv x_v$ and two others for loop densities; next we drop primes.

It should be noted that a formation of point defects is of thermo-fluctuating
character and the corresponding probability increases with a growth in the
temperature, irradiation or defects density \cite{mirz1}. As was shown in
Ref.\cite{mirz2} physically it relates to a change in the activation barrier
for defects formation due to elastic deformation of the medium caused by
defects presence. To take into account such effects one can introduce the
component relevant to this mechanism in the form: $G\exp(E_e(r)/T)$, where the
prefactor $G=p\omega_D\gamma/\lambda_v e^{-(E^f_v+E^m_v)/T}$ describes
probability of this process governed by vacancy formation and migration
energies $E^f_v$ and $E^m_v$, Debye frequency $\omega_D$ and the probability
factor $p\ll 1$; $E^e(r)$ is the activation energy changing due to elastic
field produced by defects. Following Ref.\cite{mirz2}, we assume
$G\exp(\varepsilon x/(1+x^2))$, where $\varepsilon\equiv 2ZE_e^0/T$ is given by
the energy of deformations $E_e^0$ and coordination number $Z$. Generally this
additional term is essential at laser irradiation comparing to defects
production in cascades. Next, following Ref.\cite{EPJB2012} we retain this term
without loss of generality.

From physical viewpoint vacancies are mobile species. Therefore, we have to
introduce their flux. Generally, it contains pure diffusion part $-L_d^2\nabla
x$ with diffusion length $L_d^2\equiv D_v/\lambda_v$ and the component
describing defects interaction $\mathbf{v}x =-(L_d^2/T)x\nabla U$. For the
interaction potential we assume self-consistency relation
\cite{BHKM97,PhysicaD,MetFiz09,EPJB2012} $U=-\int \tilde u(r,r')x(r'){\rm
d}r'$, where $-\tilde u(r)$ is the attraction potential with properties $\int
\tilde u(r) r^{2n+1}{\rm d}r=0$. Assuming that $x(r)$ does not change
essentially on the distance $r_0\simeq \Omega^{1/3}$, one can use an expansion
\begin{equation}\label{expansionU}
\frac{1}{T}\int{\rm d}\mathbf{r}' \tilde
u(\mathbf{r}-\mathbf{r}')x(\mathbf{r}')\simeq \varepsilon (x+ r_0^2\nabla^2x).
\end{equation}
Here the first term leads to the well-known relation between the elastic field
potential and the defects concentration $U=-\kappa\varpi \nabla\cdot
\mathbf{u}$; for the displacement vector $\mathbf{u}$ one has
$\nabla\cdot\mathbf{u}\propto\varpi x$, $\kappa$ is the elastic constant,
$\varpi$ is the dilatation parameter \cite{mirz1}. The second part in
Eq.(\ref{expansionU}) is responsible for microscopic processes of defect
interactions in the vicinity of the interaction radius $r_0$. Under normal
conditions this term is negligible comparing to the ordinary diffusion one.
However, in the absence of the second term in Eq.(\ref{expansionU}) for the
flux one gets $\mathbf{J}\propto -(1-\varpi\kappa x/T)\nabla x$, where the the
concentration depending diffusion coefficient $(1-\varpi\kappa x/T)$ can be
negative at some interval for $x$. It means that homogeneous distribution of
defects starting from some critical speed of its formation related to the
temperature, sinks density, and dilatation volume becomes unstable. The
emergence of the directional flux of defects results in supersaturation of
vacancies and formation of clusters or pores. From mathematical viewpoint such
divergence appeared at short time scales can not be compensated by nonlinear
part of the reaction terms. The second term in expansion (\ref{expansionU}) can
prevent divergencies of the derived model due to microscopic properties of
defect interactions. Therefore, the term with the second derivative must be
retained. It will be shown that this term governs the typical sizes of defect
clusters.

Taking into account all above mechanisms one arrives at the system of dynamical
equations of the form
\begin{equation}\label{le16}
\begin{split}
\partial_{t}x=&P(1-\varepsilon_v)-(1+\rho_i+\rho_v)
(x-x_0)\\&-\frac{P\epsilon(1-\varepsilon_i)x}{A(1+\rho_i+\rho_v)+{\epsilon}
x}+G e^{\frac{\varepsilon x}{1+x^{2}}}-\nabla\cdot\mathbf{J};\\
 \tau_i\partial_{t}\rho_i&=\varepsilon_iP+\frac{\tilde A A P(1-\varepsilon_i)}{A(1+\rho_i+\rho_v)
  +{\epsilon}x}- (x-x_{0});\\
 \tau_v\partial_{t}\rho_v&=\varepsilon_v P-\rho_v\left(\frac{\tilde A A P(1-\varepsilon_i)}{A(1+\rho_i+\rho_v)
  +{\epsilon}x}-(x-x_{0})\right),
\end{split}
\end{equation}
where the renormalized defect flux is
\begin{equation}
\mathbf{J}\equiv -\left[
 \nabla x -\varepsilon x\nabla(x+\ell^2\nabla^2x )\right].
\end{equation}
Here we have used renormalization $\mathbf{r}'=\mathbf{r}/L_d$ and introduced
dimensionless length $\ell^2=r_0^2/L_d^2$ with time scales $\tau_i\equiv b
r_v^0\rho_N\gamma$, $\tau_v\equiv b\alpha\rho_N/2\pi ND_v$; $\Delta B=B-B'$,
$A\equiv 1+B$, $\tilde A\equiv 1+\Delta B$.

The time scale for the vacancy population evolution is several times larger
than relaxation of cascades. It means that a particle bombardment occurs every
time at new spatially organized system. In other words, the cross-section of
defects formation or defect damage rate and/or cascade collapse efficiency
generally can be considered as fluctuating parameters
\cite{Kiritani,Kiritani_MCF97}. In such a case, formally, one can consider
stochastic nature of $P$ assuming $P\to \langle P\rangle+\zeta(\mathbf{r},t)$
where $\zeta(\mathbf{r},t)$ represents corresponding fluctuations with
properties $\langle\zeta\rangle=0$ and
$\langle\zeta(\mathbf{r},t)\zeta(\mathbf{r}',t')\rangle=2\sigma^2\delta(\mathbf{r}-\mathbf{r}',t-t')$.
Here $\sigma^2$ is the noise intensity proportional to $P$ meaning that
fluctuating source emerges only if irradiation is included; $\delta$-correlated
noise describes fast relaxation of cascades comparing to evolution of the
population field. In such a case this term should be included into dynamics of
the system (\ref{le16}) where stochastic dynamics we will treat in the
Stratonovich calculus. As a reference system we consider pure nickel with
following material constants: $E^f_v=1.8eV$, $E^m_v=1.04~eV$, $E^m_i=0.3~eV$,
 $E^e_0=0.01\div0.2~eV$, $D_v=6\cdot10^{-5}e^{-E^m_v/T}~m^2/sec$,
 $D_i=10^{-7}e^{-E^m_i/T}~m^2/sec$, $c_{0v}=e^{-E^f_v/T}$,   $\omega_D=1.11\cdot 10^{13}~sec^{-1}$
 $r^0_v=1.5\cdot 10^{-9}~m$, $\varepsilon_v=0.1$, $\varepsilon_i=0.01$,
 $\rho_N=10^{14}~m^{-2}$,  $\Omega=1.206\cdot10^{-29}~m^3$.

\section{One-component system}

Comparing time scales for vacancy and interstitial loop densities with time
scale for vacancy population one finds that $\tau_i\gg\tau_v\gg 1$. It allows
one to consider $\rho_i$ and $\rho_v$ as slow quantities. Hence, we can
consider a spatial modulation of the vacancy concentration field assuming
$\mu\equiv(1+\rho_i+\rho_v)\approx const$. Therefore, the system (\ref{le16})
can be effectively reduced to the one-component model with the external noise:
\begin{equation}\label{le16X}
\partial_{t}x=R_x(x)-\nabla\cdot\mathbf{J}+g_x(x)\zeta(\mathbf{r},t),
\end{equation}
where
\begin{equation}\label{Rxgx}
\begin{split}
 &R_x(x)=P(1-\varepsilon_v)-\mu
(x-x_0)-\frac{P\epsilon(1-\varepsilon_i)x}{A\mu+{\epsilon} x}+G
e^{\frac{\varepsilon x}{1+x^{2}}},\\
 &g_x^2(x)=(1-\varepsilon_v)^2+\left(\frac{\epsilon(1-\varepsilon_i)x}{A\mu+{\epsilon}
 x}\right)^2.
\end{split}
\end{equation}

\begin{figure}
\centering
\includegraphics[width=0.95\columnwidth]{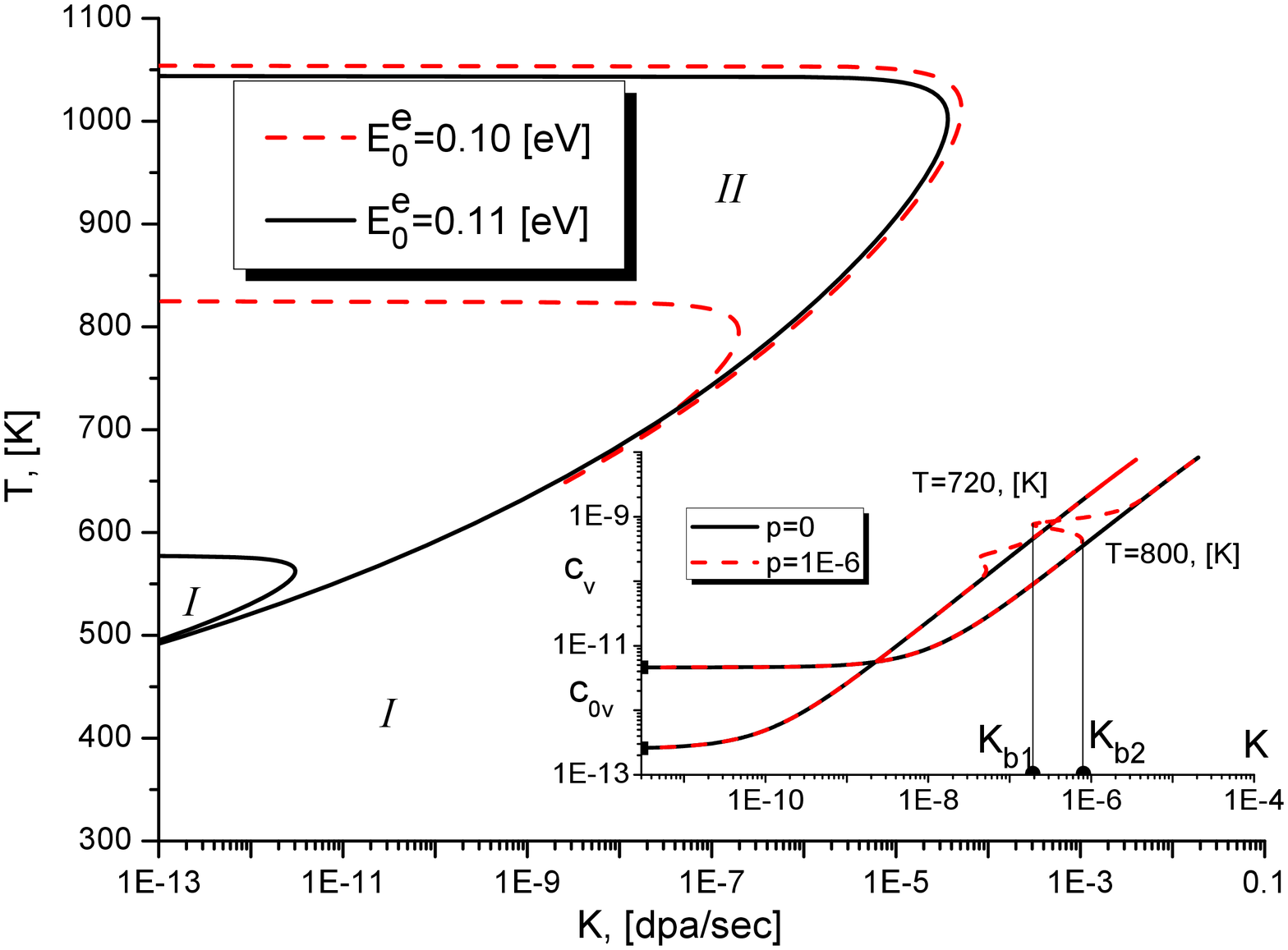}
\caption{Phase diagram and stationary states of the deterministic
system\label{fig1}}
\end{figure}

Let us consider stability of the stationary homogeneous state $x_{s}$ as a
solution of the equation $R_x(x)=0$. The corresponding phase diagram in the
plane $(K,T)$ and stationary dependencies of vacancy concentration are shown in
Fig.\ref{fig1}. It follows that due to influence of the elastic field the
system manifests bistable regime (cf. curves in insertion at different $p$). In
the domains $I$ the system is monostable. The bistability domain $II$ bounded
by interval $K\in[K_{b1},K_{b2}]$ shrinks with decrease in $E^0_e$.

\begin{figure}[!ht]
\centering
\includegraphics[width=0.95\columnwidth]{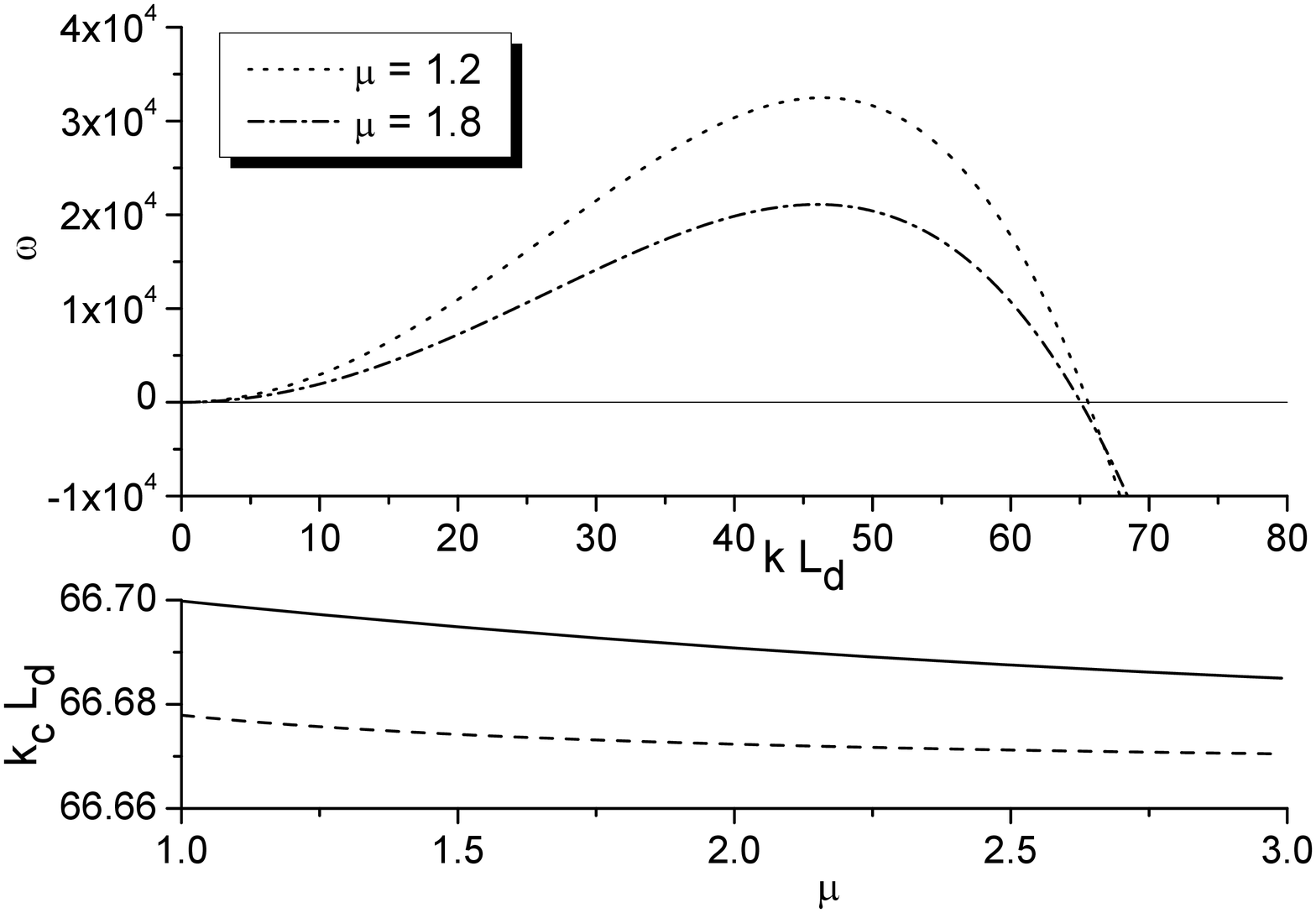}
\caption{Dispersion law at different sink densities under reactor conditions
(top) and dependencies of critical wave number \emph{versus} sink densities
(bottom) under accelerator (solid line) and reactor (dash line)
conditions\label{omega(k)}}
\end{figure}

Considering dynamics of the averaged fluctuation $\langle \delta
x\rangle=\langle x\rangle-x_s$ in the Fourier space we obtain
\begin{equation}
\frac{{\rm d}\langle \delta x\rangle}{{\rm d}t}= \left(\Lambda
+\omega(k)\right) \langle \delta x\rangle,
\end{equation}
where the Lyapunov exponent is
\begin{equation}
\begin{split}
\Lambda&=-\mu-{\frac {{P}(1-\varepsilon_i)\epsilon\,\mu\,A}{\left
[A\mu+x_s\epsilon\right ]^{2}}}
+G\frac{\varepsilon(1-x_s^2)}{(1+x_s^2)^2}\exp\left(\frac{\varepsilon x_s
}{1+x_s^2}\right)\\
&+\frac{\sigma^2\epsilon^2(1-\varepsilon_i)^2}{\left[A\mu+\epsilon
x_s\right]^2}\left(1-\frac{\epsilon x_s}{A\mu+\epsilon
x_s}\left[4-\frac{3\epsilon x_s}{A\mu+\epsilon x_s}\right]\right).
\end{split}
\end{equation}
Here the last term represents a contribution from the Stratonovich drift
leading to destabilization of the homogeneous state. The noise action leads to
destabilization increasing the value of $\Lambda$. A dispersion relation
describing instability with respect to inhomogeneous perturbations is given by
\begin{equation}\label{eqw(k)}
\omega(k)=-k^2[1-\varepsilon x_s(1-\ell^2k^2)].
\end{equation}
It follows that unstable modes with $\omega(k)>0$ are characterized by
wave-numbers $0<k<k_c$, where
\begin{equation}\label{k_c_det}
 k_c=\sqrt{\frac{\varepsilon x_s-1}{\varepsilon x_s
 \ell^2}}.
\end{equation}
It is seen that in the simplest case of $\ell\to 0$ in monostable domain (below
the cusp) all states with $x_s>1/\varepsilon$ are unstable with respect to
inhomogeneous perturbations with $k_c\to\infty$, whereas states with
$x_s<1/\varepsilon$ are stable. In the actual case $\ell\ne0$ the system states
characterized by $x_s>1/\varepsilon$ are unstable with wave numbers lying in
the interval $0<k<k_c$. In the bimodal domain the system is always unstable
with respect to inhomogeneous perturbations. The wave number for the most
unstable mode $k_0$ can be found from the solution of the equation ${\rm
d}\omega(k)/{\rm d}k=0$ giving $k_0=k_c/\sqrt{2}$. Dependencies $\omega(k)$ at
different values for $\mu$ are shown in Fig.\ref{omega(k)} (top panel). An
increase in sinks density leads to lower value of $x_s$. This results to a
stabilization of the uniform defects distribution, as was shown previously in
Ref.\cite{Walgraef95}: in the studied system the maximal value of $\omega(k)$
decreases resulting to a shift of $k_c$ toward small values (see
Fig.\ref{omega(k)} (bottom panel)). Hence, at elevated sink densities a number
of unstable modes decreases and period of patterns characterized by most
unstable mode $k_0$ grows.

Next, let us consider the system behavior in the stationary case setting
$\ell\lll 1$. It allows one to use mean-field approximation resulting in
$\nabla^2x\to (x-\eta)$ and $(\nabla x)^2\to(x-\eta)^2$, where
$\eta\equiv\langle x\rangle$ is the mean field (in systems undergoing
order-disorder phase transitions it plays a role of the order parameter)
describing ordering effects \cite{Garcia}. The averaging is provided according
to stationary distribution function
$\mathcal{P}_s(t\to\infty;x,\eta)=\mathcal{P}_s(x,\eta)$ as solution of the
corresponding Fokker-Planck equation \cite{Risken}:
\begin{widetext}
\begin{equation}
\frac{\partial\mathcal{P}}{\partial t}=-\frac{\partial}{\partial x
}\left(R_x(x)+(1-\varepsilon
x)(x-\eta)-\varepsilon(x-\eta)^2\right)\mathcal{P}+\sigma^2\frac{\partial}{\partial
x}g_x(x)\frac{\partial}{\partial x}g_x(x)\mathcal{P}.
\end{equation}
Its stationary solution
\begin{equation}
\mathcal{P}_s(x;\eta)=\frac{1}{\mathcal{N}(\eta)}\exp\left(\frac{1}{\sigma^2}\int\limits_0^x\frac{R_x(y)-(1-\varepsilon
y)(y-\eta)+\varepsilon(y-\eta)^2-\sigma^2 g_x(y)g_x'(y) }{g_x^2(y)}{\rm d}y
\right)
\end{equation}
\end{widetext}
defines the mean field in a self-consistent manner
\begin{equation}\label{eq_sc}
\eta=\int\limits_0^\infty x\mathcal{P}_s(x;\eta){\rm d}x;
\end{equation}
constant $\mathcal{N}(\eta)$ takes care of the normalization condition
$\int_0^\infty\mathcal{P}_s(x;\eta){\rm d}x=1$.

\begin{figure}
\centering a)
\includegraphics[width=0.95\columnwidth]{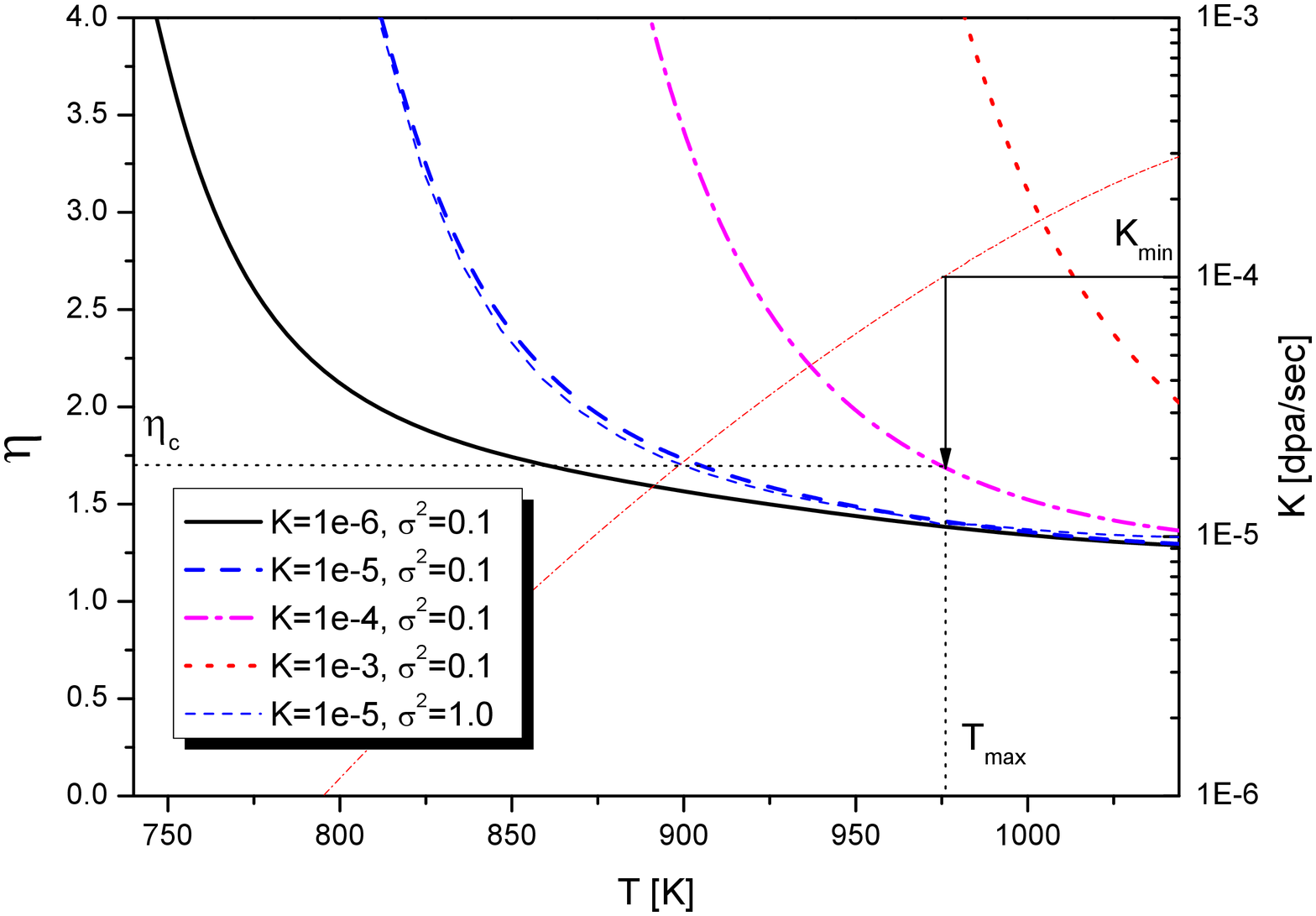}\\
b)
\includegraphics[width=0.95\columnwidth]{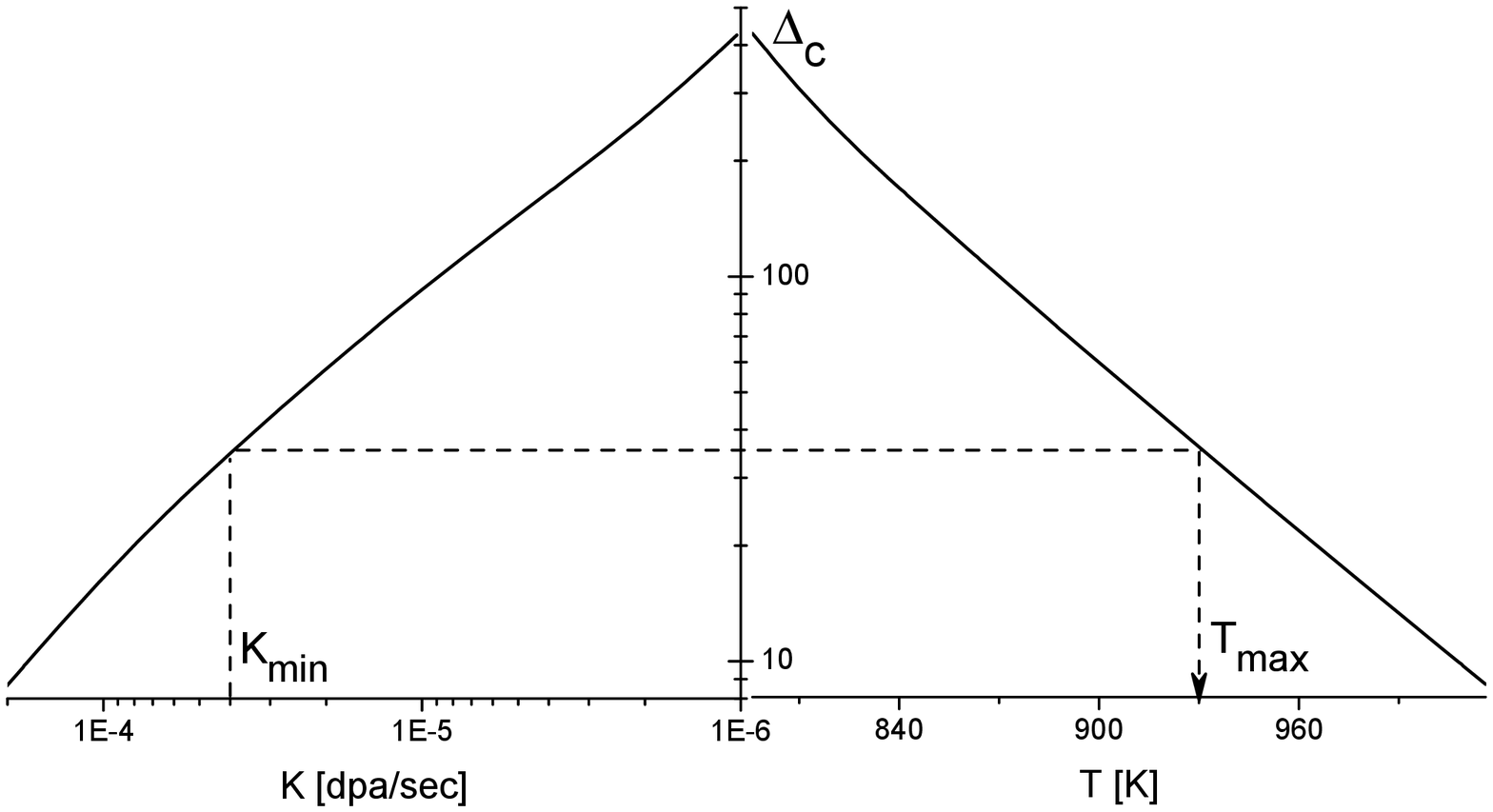}
\caption{Mean field results: a) mean field dependence on the temperature at
different defect damage rates and the external noise intensity (a procedure to
determine critical value $\eta_c$ is shown by arrows with the help of the
linear stability diagram in the plane $(T,K)$); b) two-dimensional plot of the
critical supersaturation dependence
$\Delta_c\equiv\Delta_c(K,T)=(\eta_c(K,T)-x_0(T))/x_0(T)$ at
$\sigma^2=0.1$\label{mf}}
\end{figure}

Solutions of the self-consistency equation (\ref{eq_sc}) are shown in
Fig.\ref{mf}a. It follows that the mean field falls down with the temperature
increase meaning a loss of spatial order. Comparing the corresponding curves at
different defect damage rates one obtains that spatial organization is larger
at elevated $K$ due to formation of new interacting defects. The external noise
action does not change principally the dependencies $\eta(K,T)$. Here one has
small decrease of the mean field at actual temperature regime relevant to
reactors, for example, and small increase in $\eta$ at elevated temperatures.
As far as external fluctuations decrease $\eta$ the stationary value of
concentration $x_s$ (in stochastic case it is reduced to most probable value)
decreases too leading to an extension of the number of unstable modes,
comparing to the deterministic case (see Eqs.(\ref{eqw(k)},\ref{k_c_det})).
Therefore, the period of pattern described by the wave number $k_0$ grows with
the noise influence.

Using mean field results one can obtain critical values of $\eta$ limiting
supersaturation of point defects above which spatial modulation is realized.
Indeed, as far as $\eta$ is a measure of the averaged point defects population
$\eta\equiv \langle x\rangle$ we can define the supersaturation in the standard
manner $\Delta\equiv(\eta-x_0)/x_0$, where $\Delta=\Delta(K,T)$, $x_0$ relates
to the equilibrium vacancy population. Next, choosing a value of $K$ we can
obtain the critical value of the temperature from the phase diagram obtained
from the linear stability analysis (follow the arrow in Fig.\ref{mf}a). This is
the maximal temperature $T_{max}$ bounding the temperature domain $(T<T_{max})$
for spatial instability, whereas $K_{min}$ is the minimal damage rate
$(K>K_{min})$ bounding spatial instability. The critical value $\eta_c$ we
define as $\eta_c=\eta(K_{min}, T_{max})$. Therefore, we arrive at the
dependence of the critical supersaturation $\Delta_c(K,T)$ of point defects
limiting spatial organization of defect structure (see Fig.\ref{mf}b). It
follows that $\Delta_c$ decreases with growth of both $K$ and $T$ meaning
emergence of ordering/patterning processes at low population of point defects
at elevated $K$ or $T$. The physical reason for this effect lies in production
of large amount of defects by irradiation and their generation by the
thermo-activation mechanism.

\section{Three-component system}
In the three-component stochastic model we take $\rho_{i,v}=\rho_{i,v}(t)$,
giving $\mu=\mu(t)$. Therefore, instead of the deterministic system
(\ref{le16}) we arrive at
\begin{equation}
\begin{split}
\partial_{t}x&=R_x(x;\rho_i, \rho_v)-\nabla\cdot\mathbf{J}+ g_x(x;\rho_i, \rho_v)\zeta(\mathbf{r},t);\\
 \tau_i\partial_{t}\rho_i&=R_i(\rho_i;x, \rho_v)+g_i(\rho_i; x, \rho_v)\zeta(\mathbf{r},t);\\
 \tau_v\partial_{t}\rho_v&=R_v(\rho_v; x,\rho_i)+g_v(\rho_v; x,\rho_i)\zeta(\mathbf{r},t),
\end{split}
\end{equation}
here $R_x$ and $g_x$ are given by Eq.(\ref{Rxgx}). For the deterministic
components $R_i$, $R_v$ and noise amplitudes $g_i$, $g_v$ one has:
\begin{equation}
\begin{split}
R_i&\equiv \varepsilon_iP+\frac{\tilde A A
P(1-\varepsilon_i)}{A(1+\rho_i+\rho_v)
  +{\epsilon}x}- (x-x_{0}),\\
R_v&\equiv \varepsilon_v P-\rho_v\left(\frac{\tilde A A
P(1-\varepsilon_i)}{A(1+\rho_i+\rho_v)
  +{\epsilon}x}-(x-x_{0})\right),\\
g^2_i&\equiv \varepsilon_i^2+\left(\frac{\tilde A A
(1-\varepsilon_i)}{A(1+\rho_i+\rho_v)
  +{\epsilon}x}\right)^2,\\
g^2_v&\equiv \varepsilon_v^2+\left(\frac{\tilde A A
(1-\varepsilon_i)\rho_v}{A(1+\rho_i+\rho_v)
  +{\epsilon}x}\right)^2.
 \end{split}
\end{equation}

In our study we make numerical analysis on 2-dimensional square lattice
$N\times N$ where $N=128$ with periodic boundary conditions and mesh size
$\Delta l=1.0$. For the numerical integration the Milshtein method was used
\cite{Garcia}. The time step is $\Delta t=2.5\times 10^{-4}$. In all numerical
experiments initial conditions correspond to equilibrium concentration of
vacancies with no loops, i.e, $\langle c(\mathbf{r},0)\rangle=c_{eq}$, with
small dispersion $\langle(\delta c(\mathbf{r},0))^2\rangle=10^{-1}c_{eq}$;
$\langle \rho_i(\mathbf{r},0)\rangle=\langle \rho_v(\mathbf{r},0)\rangle=0$. We
consider two regimes of irradiation corresponding to reactor and accelerator
conditions. In the modeling scheme we put $\ell=0.25$ in order to stabilize the
computational algorithm. An estimation for this parameter for pure $Ni$ gives
$\ell_{Ni}=0.015$. Hence, in numerical data analysis we will use the
renormalization parameter $\chi\equiv\ell_{Ni}/\ell=0.06$ to estimate a linear
size of defect clusters $\langle d_0\rangle$ and a characteristic length of
their spatial distribution $L_0$.

\subsection{Patterning under  reactor conditions}
\begin{figure}
\centering
\includegraphics[width=0.95\columnwidth]{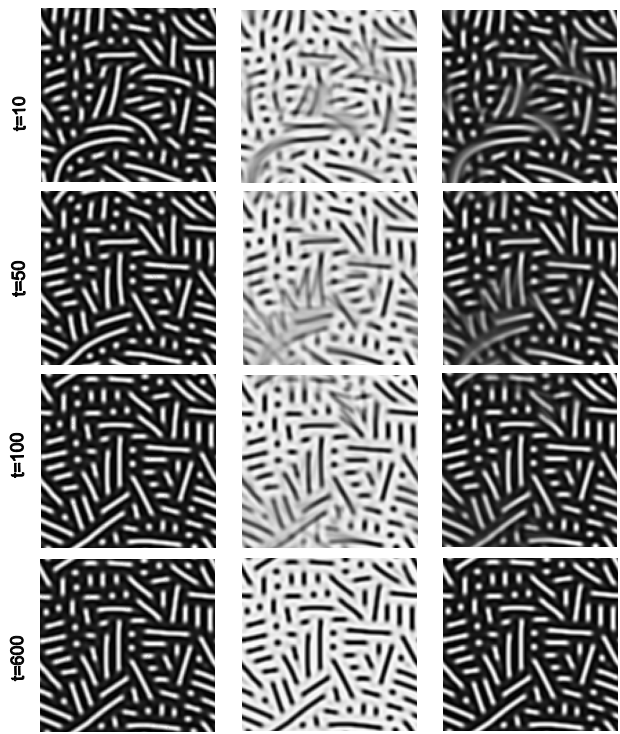}
\caption{Typical evolution of the deterministic ($\sigma^2=0$) system under
reactor conditions ($T=773$K, $K=10^{-6}~dpa/sec$). Columns (from the left to
the right) denote dynamics of fields $x$, $\rho_i$,
$\rho_v$.\label{evol_reactor}}
\end{figure}

Let us consider patterning under reactor conditions. Typical snapshots of the
system evolution are shown in Fig.\ref{evol_reactor}. Here first, second and
third columns correspond to dynamics of the dimensionless fields
$x(\mathbf{r},t)$, $\rho_i(\mathbf{r},t)$ and $\rho_v(\mathbf{r},t)$,
respectively. It is seen that at early stages of the system evolution starting
from the Gaussian distribution of vacancies around the equilibrium value the
spatial arrangement of defects is realized. Here small spherical clusters and
extended ones representing defect walls are organized (see first column). Their
spatial structure is slightly changed at further system evolution. Increasing
the irradiation dose $Kt$ one can observe well organized structure of vacancy
loops repeating spatial structure of vacancy clusters, whereas interstitial
loops density behaves vice verse to the vacancy loops density. Moreover, at
early stages one can find smearing of the fields $\rho_i$ and $\rho_v$ in the
vicinity of the extended defects. With the irradiation dose increase the number
of loops grows.

\begin{figure}
\centering
\includegraphics[width=0.95\columnwidth]{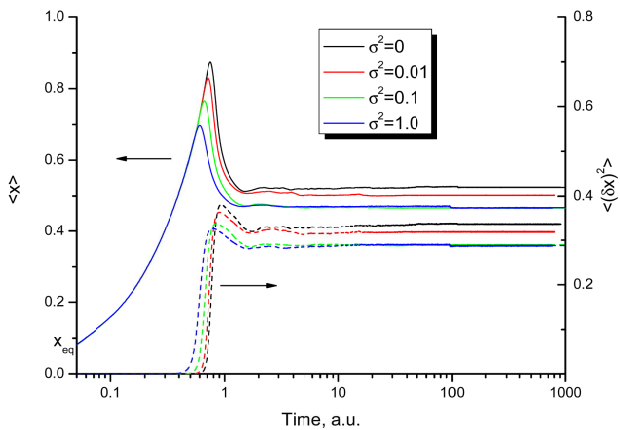}
\caption{Dynamics of the averaged concentration field and its variance
$\langle(\delta x)^2\rangle$ at different values for the external noise
intensity $\sigma^2$ under reactor conditions.\label{evol_aver}}
\end{figure}

\begin{figure} \centering
 a)\includegraphics[width=0.95\columnwidth]{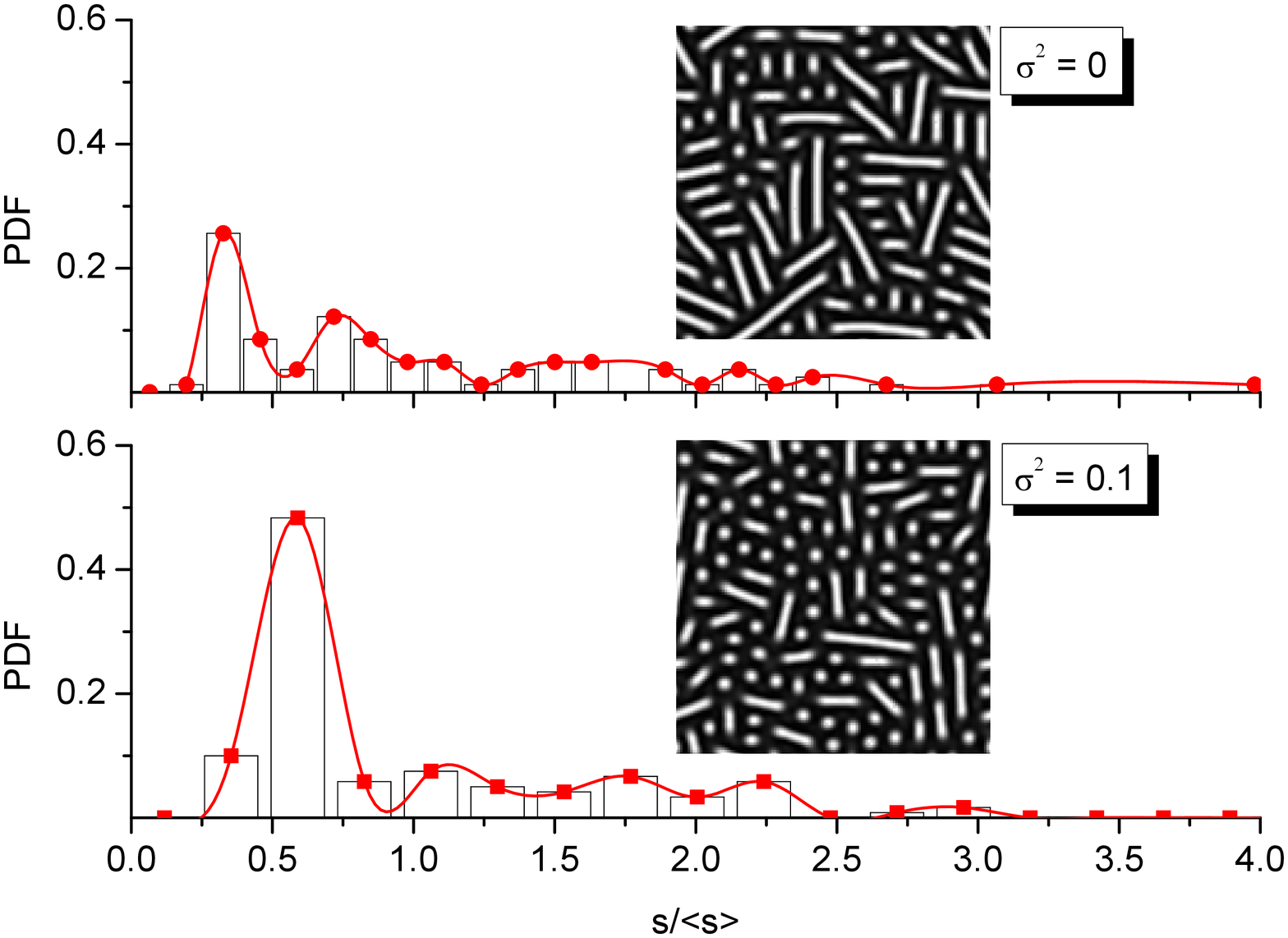} \
 b)\includegraphics[width=0.95\columnwidth]{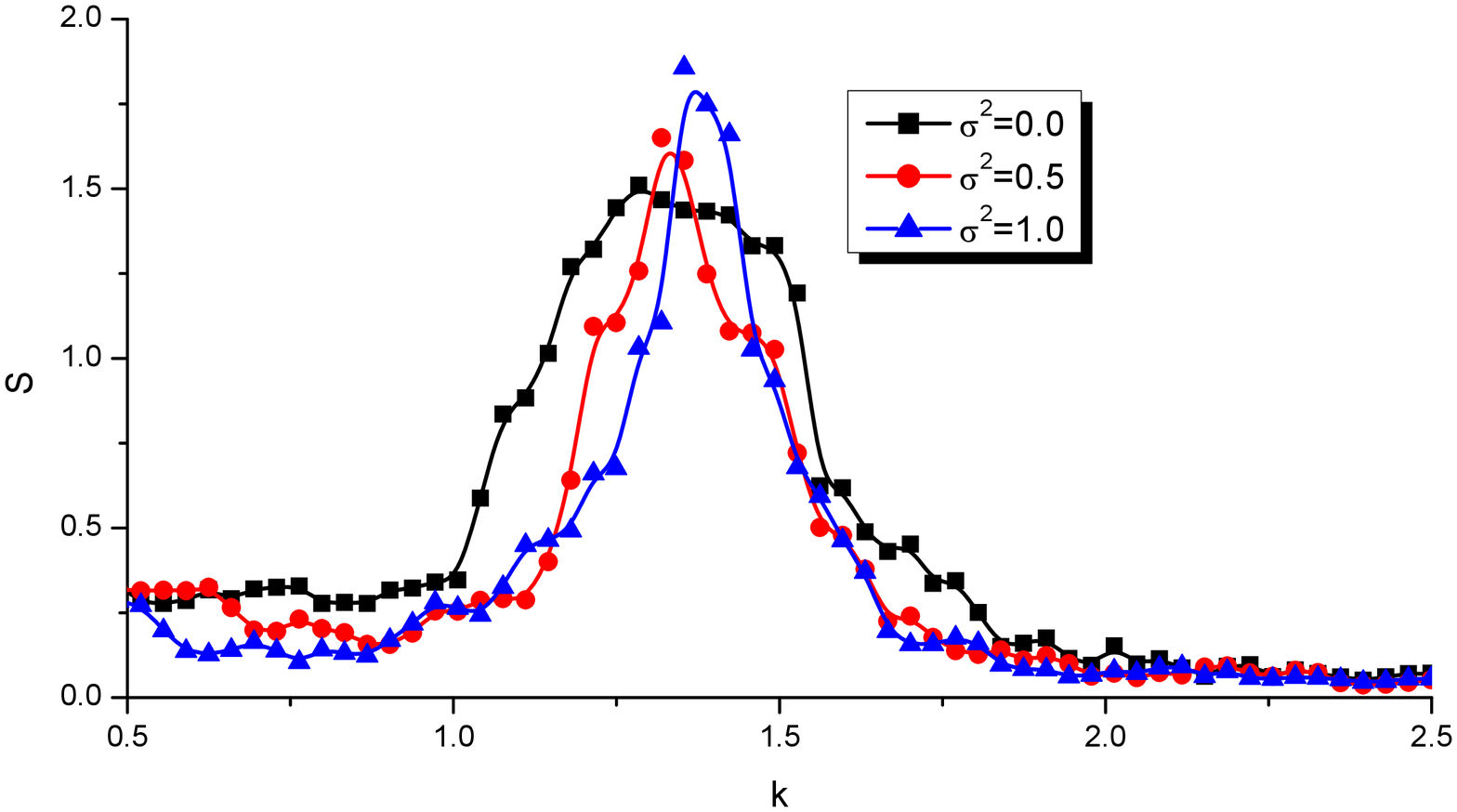}
 c)\includegraphics[width=0.95\columnwidth]{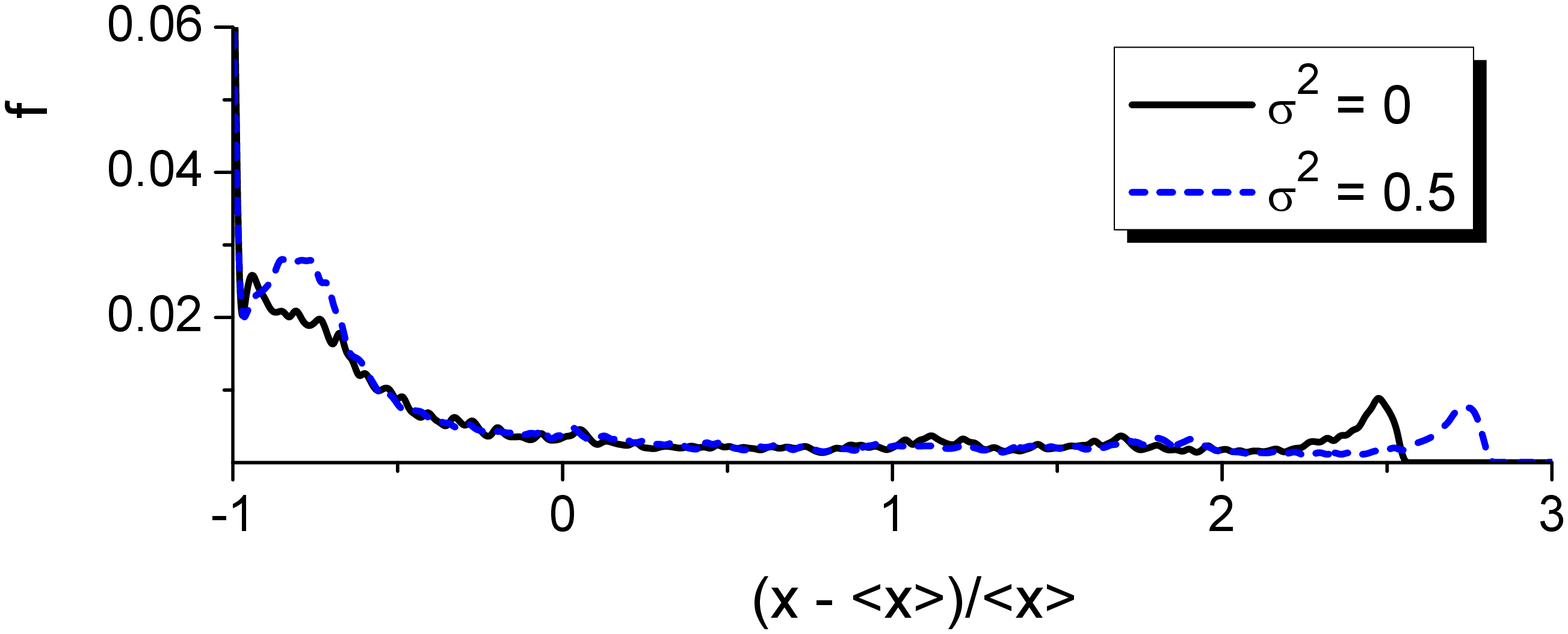} \caption{(a)
Stationary probability density for vacancy clusters distribution over sizes $S$
for the deterministic ($\sigma^2=0$) and stochastic ($\sigma^2=0.5$) cases. (b)
Spherically averaged structure function for the vacancy concentration at
different noise intensities at $t=800$. Other parameters are: $T=773$K,
$K=10^{-6}~dpa/sec$. (c) Distribution functions over vacancy clusters occupancy
at deterministic and stochastic conditions\label{vac_noise}}
\end{figure}

To study self-organization of vacancy clusters we consider dynamics of the
quantity $\langle(\delta x)^2\rangle$. In patterning processes it plays a role
of the effective order parameter measuring symmetry breaking of initial
Gaussianly distributed field \cite{Garcia,PhysicaD,PhysA2010_1}. Considering
its dynamics together with the averaged concentration $\langle x\rangle$ one
finds that at early stages the number of point defects increases toward its
maximal value (see Fig.\ref{evol_aver}).
From this time the variance $\langle(\delta x)^2\rangle$ grows meaning
formation of spatial order. Here due to a supersaturation of vacancies produced
by irradiation cascades a vacancy enriched phase (spherical clusters and defect
walls) appears. As a result the number of free vacancies decreases due to
migration to sinks. An emergence of weak oscillations here corresponds to
processes when migrating vacancies can be absorbed by clusters and emitted from
them absorbing by sinks. At irradiation dose increase the quantity $\langle
x\rangle$ attains the stationary value with small fluctuations. It means that
new vacancies formed in cascades move to sinks with small amount in a bulk.
Domains having small amount of vacancies contain elevated interstitial loop
densities (see second column in Fig.\ref{evol_reactor}). Comparing dependencies
in Fig.\ref{evol_aver} at different noise intensities one can conclude that the
noise action accelerates formation of spatial order due to instability
described by the Stratonovich drift as was shown in the previous Section. From
the other hand it reduces the averaged vacancy concentration and a value of the
order parameter. The last means small variance of the concentration field in
the system at elevated intensity of external fluctuations. In other words
another kind of the spatial modulation of vacancy field is realized with small
value of the averaged vacancy concentration. Indeed, comparing snapshots for
vacancy field distribution at different noise intensities it is seen that
external fluctuations promote spatial rearrangement of vacancies with
preferable formation of spherical clusters (see Fig.\ref{vac_noise}a). From the
obtained histograms of the vacancy cluster size distribution it follows that in
the deterministic case there are three well-defined peaks in distributions
corresponding to small spherical clusters characterized by areas $s<\langle
s\rangle$ and extended clusters with $s\simeq \langle s\rangle$ and $s>\langle
s\rangle$. At elevated noise intensity extended clusters can not be formed due
to large fluctuations in defects production rate resulting in formation of
small clusters. Here the probability density takes large values at $s<\langle
s\rangle$; extended clusters have small probability for organization. Moreover,
with the noise intensity increase the peak of the corresponding distribution
shifts toward $s/\langle s\rangle=1$.

To study the noise influence on a period of patterns we consider a spherically
averaged structure function for the vacancy concentration field
$S(k,t)=N^{-2}\sum_{k<|\mathbf{k}|<k+\Delta k}S(\mathbf{k},t)$, where
$S(\mathbf{k},t)=\langle\delta x_{-\mathbf{k}}\delta x_{\mathbf{k}}\rangle$,
$\delta x=x-\langle x\rangle$, $x_{\mathbf{k}}=\int{\rm
d}\mathbf{r}x(\mathbf{r},t)e^{i\mathbf{kr}}$. The corresponding dependencies at
fixed time are shown in Fig.\ref{vac_noise}b. In the deterministic case the
peak of $S(k)$ is smeared due to realization of extended structures. At
elevated noise intensities the number of extended structures decreases and we
get a narrow peak with higher value for $S(k)$. It is well seen that period of
structures decreases slightly due to formation of the large number of spherical
vacancy clusters. The area under curves $S(k)$ decreases with the noise
intensity growth that well corresponds to lower values of the order parameter
$\langle(\delta x)^2\rangle$. From the obtained data and diffusion length
definition $L_d\propto \rho_N^{-1/2}\sim10^{7}~m$ one can conclude that the
period of vacancy clusters $L_0$ related to position of the main peak of the
structure function slightly varies in the interval $L_0\sim (4.8\div4.5)\chi
L_d$ with the noise intensity growth. Estimation for the averaged linear size
of vacancy clusters (diameter for spherical objects) $\langle d_0\rangle$ gives
$\langle d_0\rangle\simeq \chi L_d\sim 6~nm$. Using the obtained data one can
study vacancy distribution inside clusters. The corresponding distribution
functions $f(x)$ are shown in Fig.\ref{vac_noise}c in both deterministic and
stochastic cases. It is seen that most of clusters are characterized by
concentration less that mean value $\langle x\rangle$, however, there is well
defined peak related to vacancy clusters with concentration $x> \langle
x\rangle$. The external fluctuations increase vacancy concentration in clusters
shifting two main peaks of $f(x)$ toward large $x$. The smeared peak around
$x=0$ corresponds to state where some vacancies are distributed inside
crystalline matrix, not arranged into clusters. From physical viewpoint it
caused by vacancy diffusion processes playing here a major role.
\begin{figure} \centering
a)\includegraphics[width=0.95\columnwidth]{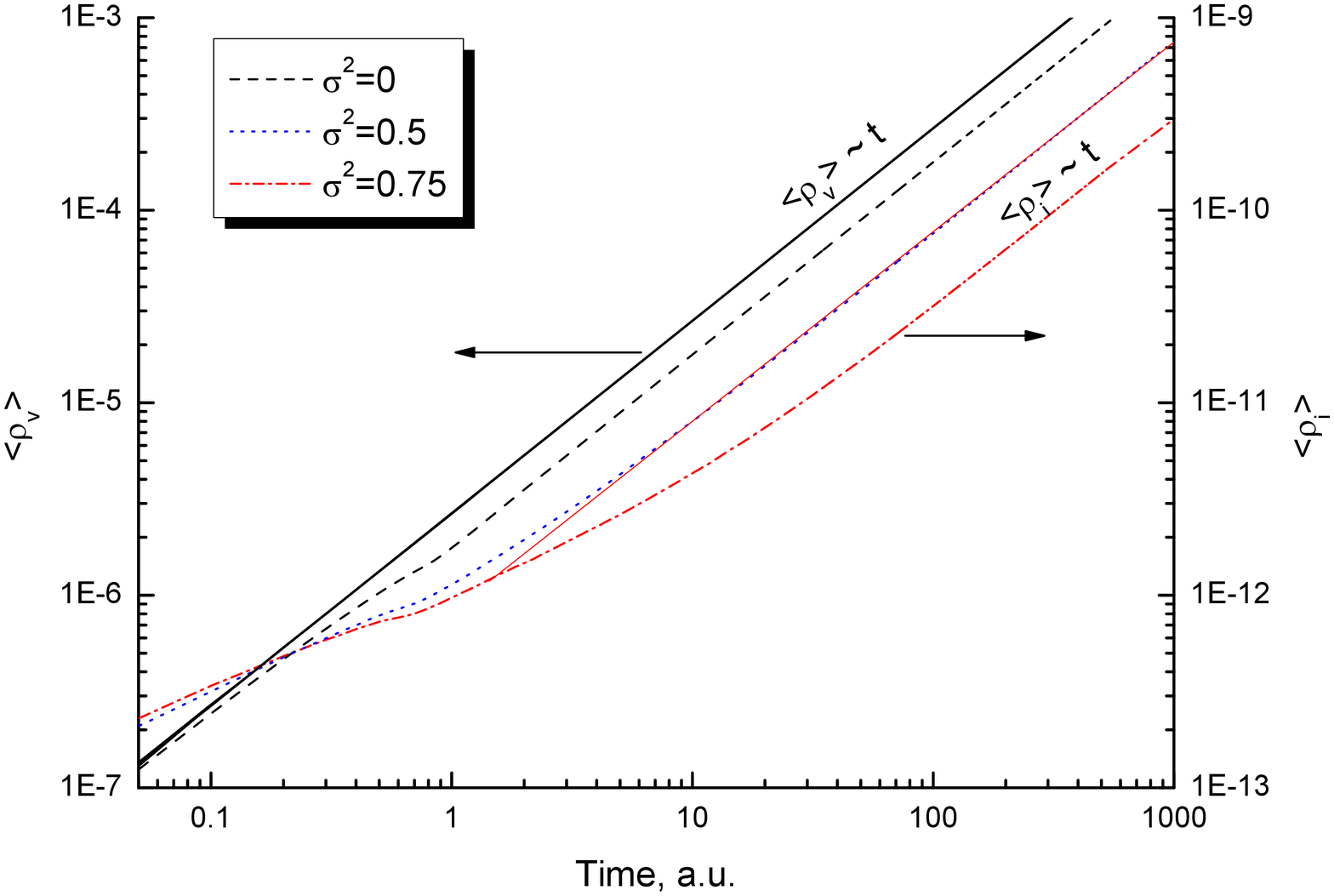} \ b)
\includegraphics[width=0.95\columnwidth]{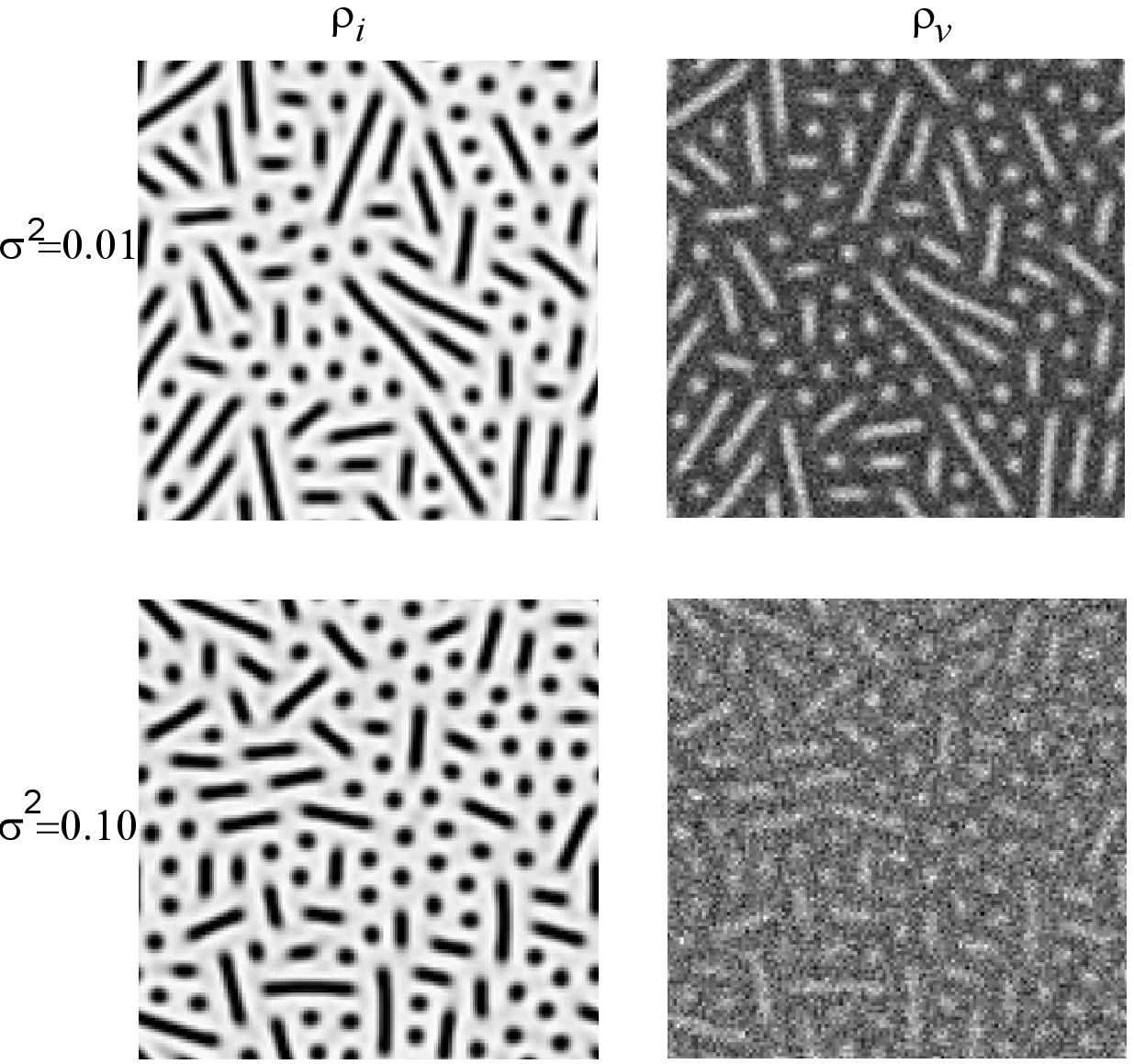}\caption{(a) Dynamics of the
averaged sink densities $\langle\rho_i\rangle$, $\langle\rho_v\rangle$ for the
different values of the external noise intensity under reactor conditions. (b)
Typical snapshots of sinks distribution at different noise intensities at
$t=700$. \label{reactor_loops}}
\end{figure}

Considering a behavior of sink densities $\langle\rho_i\rangle$,
$\langle\rho_v\rangle$ one finds that during the system evolution the
corresponding averages grow continuously with the irradiation dose increase
(see Fig.\ref{reactor_loops}a). It means that interstitial and vacancy loops
attracting vacancies grow in time due to positive feedback with vacancy field
dynamics. The external noise decreases the corresponding averages. The most
essential contribution of the noise can be seen on the sink density
$\langle\rho_i\rangle$. It is interesting to note that quantities
$\langle\rho_i\rangle$, $\langle\rho_v\rangle$ behave themselves in linear
manner at large time intervals, i.e.,
$\langle\rho_i(t)\rangle,\langle\rho_v(t)\rangle\sim t$. Such dynamics of the
sink density $\langle\rho_v\rangle$ is well observed even at early stages of
the system evolution, whereas linear growth of $\langle\rho_i\rangle$ is
realized at $t\gg 1$ only. The noise action suppresses dynamics of
$\langle\rho_i\rangle$ at initial stages moving the universal part of its
evolution toward large time scales. Such delaying dynamics leads to a weak
arrangement of sink densities. At large time intervals one has well-organized
vacancy clusters, whereas fields of sink densities have weak spatial modulation
(see snapshots of $\rho_i(\mathbf{r})$ and $\rho_v(\mathbf{r})$ at $t=1000$ at
different noise intensities in Fig.\ref{reactor_loops}b). In such a case the
distributions of fields $\rho_i(\mathbf{r})$ and $\rho_v(\mathbf{r})$ are quite
noisy.

\subsection{Patterning under  accelerator conditions}
Let us examine behavior of the system under accelerator conditions. Here we
consider the typical case of $T=900 K$ and $K=10^{-3}~dpa/sec$. In the
corresponding simulations due to large damage rate $K$ the related time step is
$\Delta t=15\times 10^{-5}$ in order to stabilize the numerical scheme.

\begin{figure} \centering
a)\includegraphics[width=0.95\columnwidth]{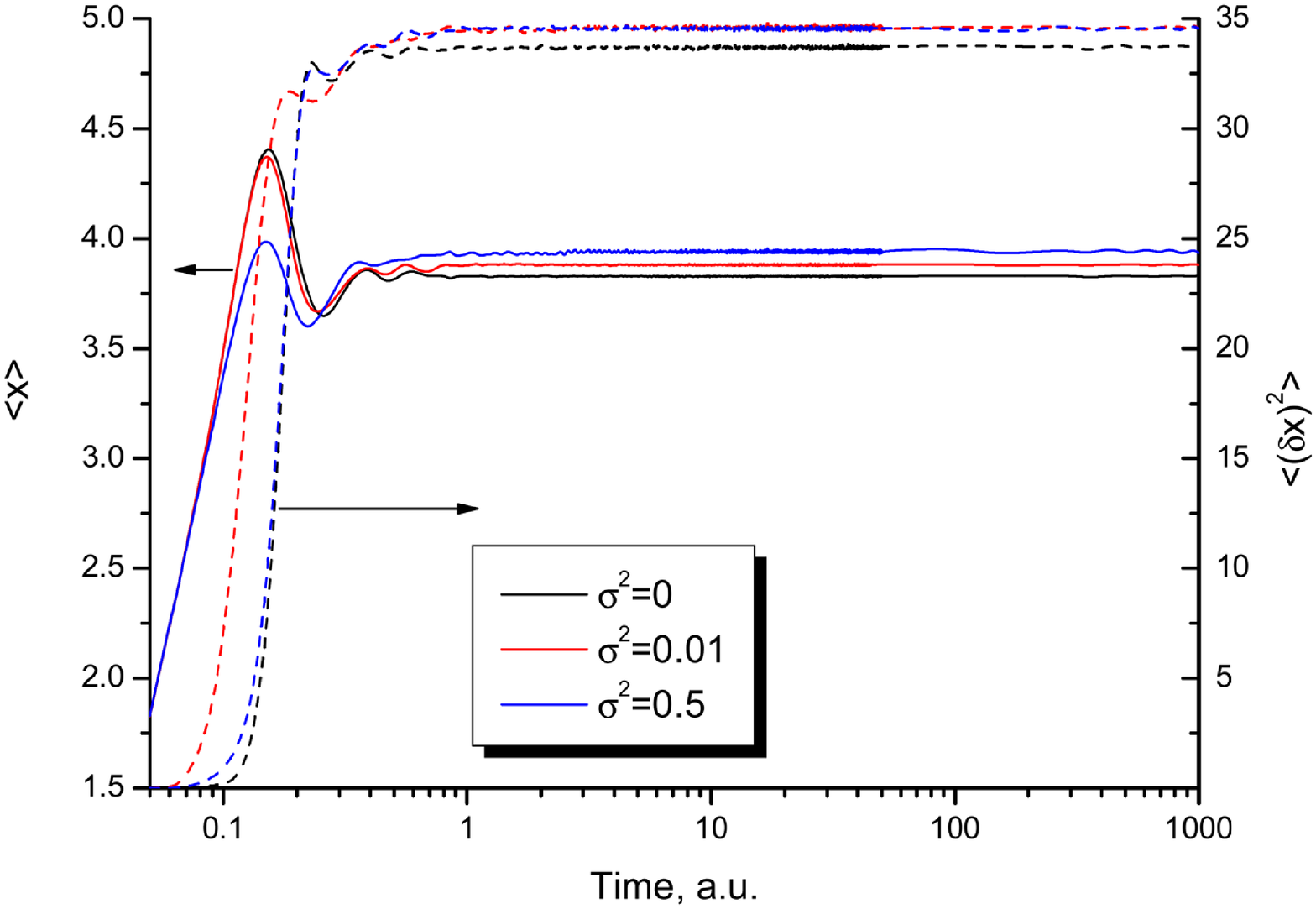} \ b)
\includegraphics[width=0.95\columnwidth]{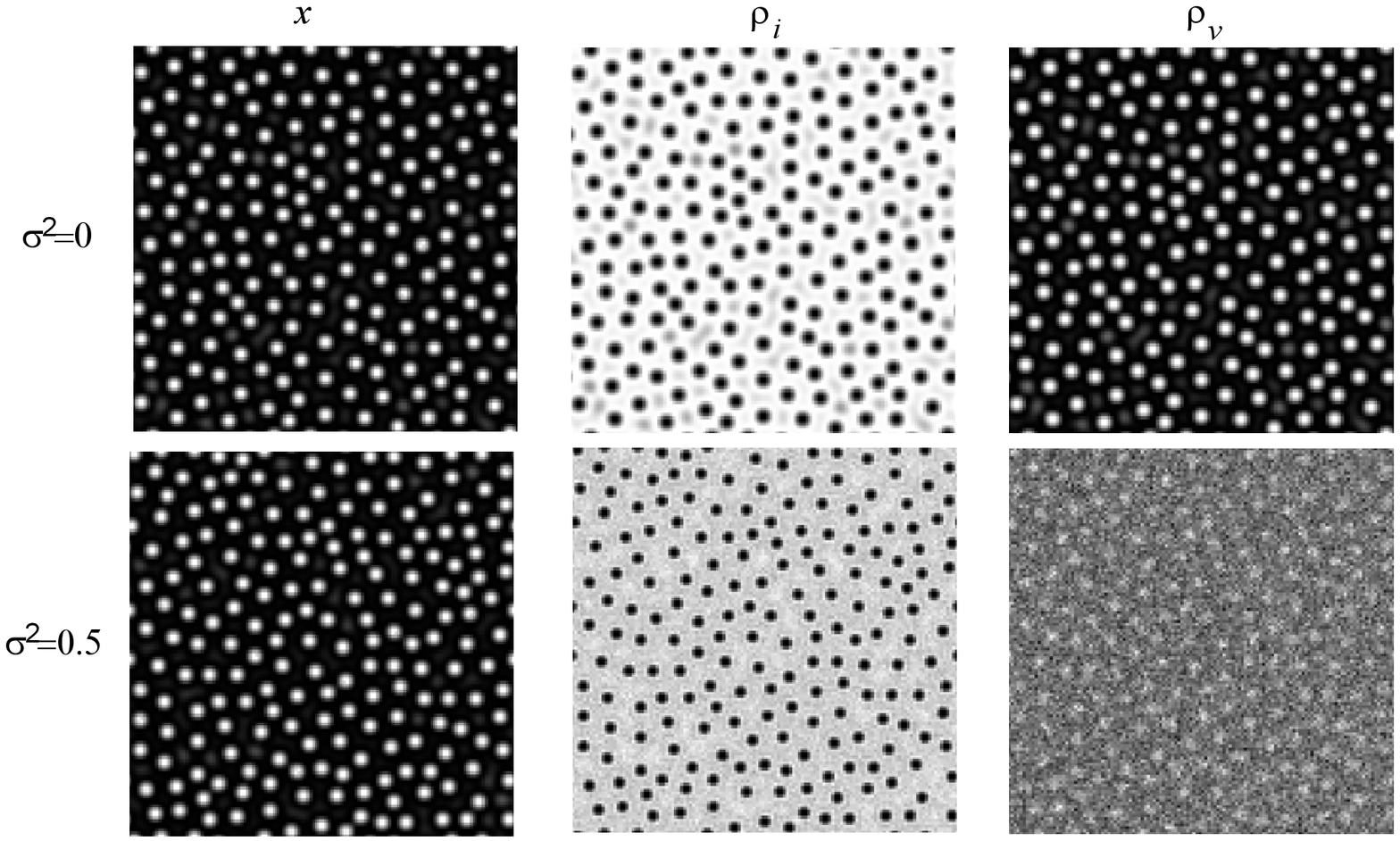}\caption{(a) Dynamics of the
averaged concentration density and the corresponding variance for different
values of the external noise intensity under accelerator. (b) Typical snapshots
of vacancy concentration field and sinks distribution in deterministic and
stochastic cases at $t=700$. \label{accelartor_x}}
\end{figure}

Comparing dynamics of the averaged vacancy concentration under reactor and
accelerator conditions one finds that in the second case the defects
arrangement occurs essentially faster (cf.Fig.\ref{evol_aver} and
Fig.\ref{accelartor_x}a). Moreover, number of defects increases by an order.
The variance $\langle(\delta x)^2\rangle$ takes large values by two orders
comparing to the previous case. It means that there is high difference in
defect concentrations over the whole system. At high damage rate defects
arrange into spherical clusters without extended structures (see
Fig.\ref{accelartor_x}b). At such conditions the noise influence is opposite to
the previous case: here due to combined effect of high-speed defect production
and fluctuation influence the number of defects increases but fluctuations
promote formation of vacancy clusters of a similar size. Considering spatial
organization of fields $\rho_i$ and $\rho_v$ it is seen that even at large
noise intensity there is well defined spatial structure of sink densities.

\begin{figure} \centering
a)\includegraphics[width=0.95\columnwidth]{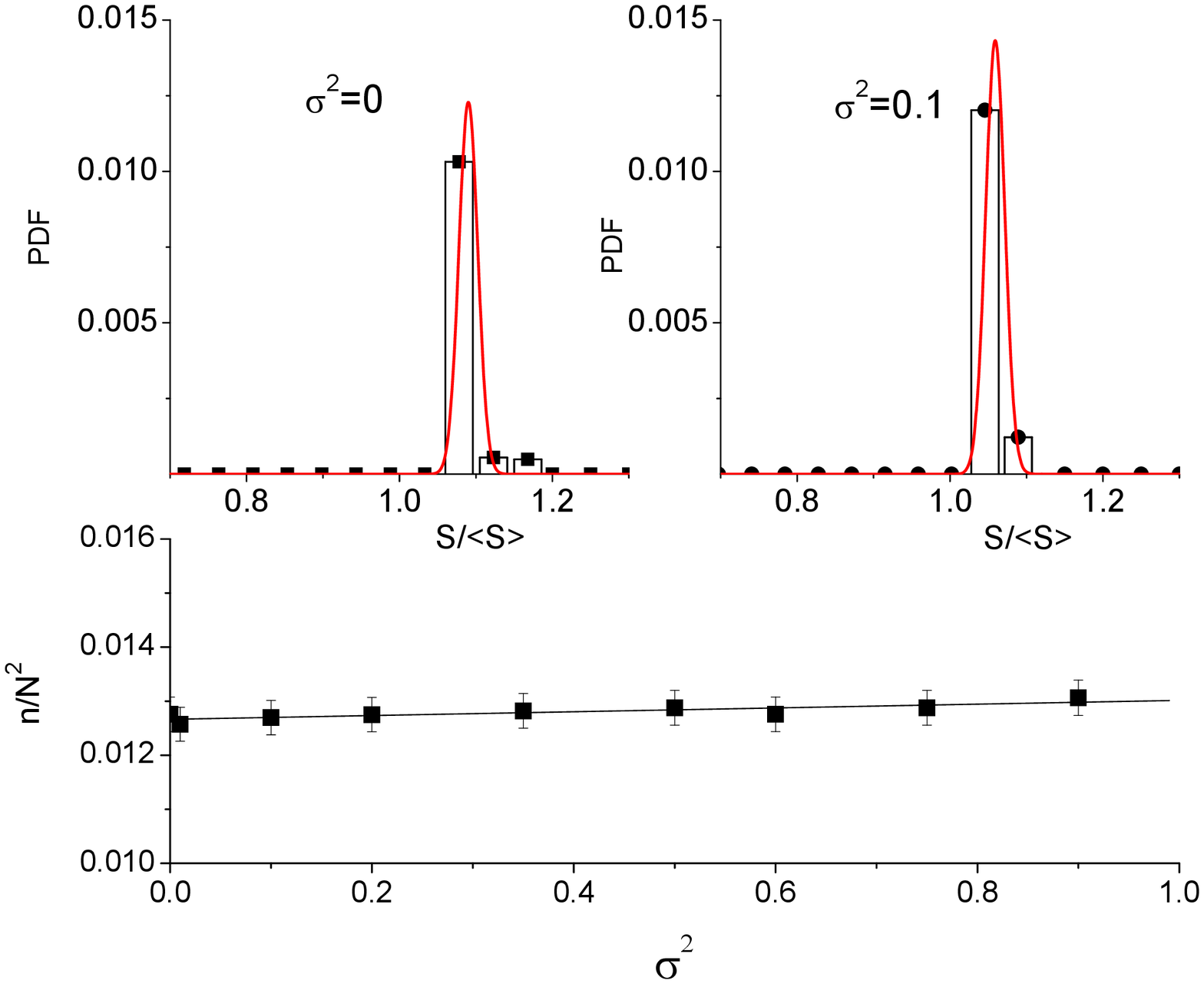}
 b)\includegraphics[width=0.95\columnwidth]{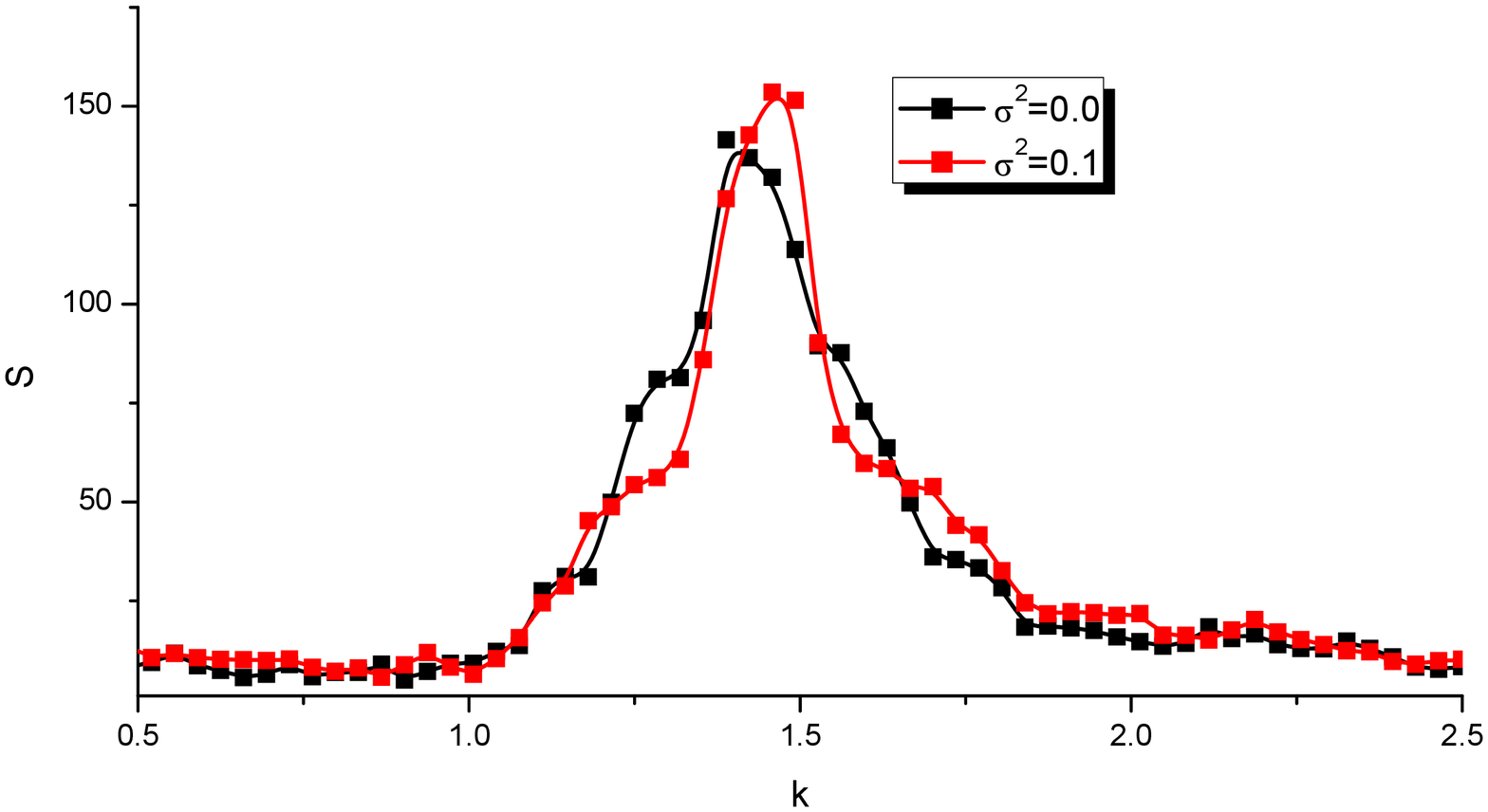}
  c)\includegraphics[width=0.95\columnwidth]{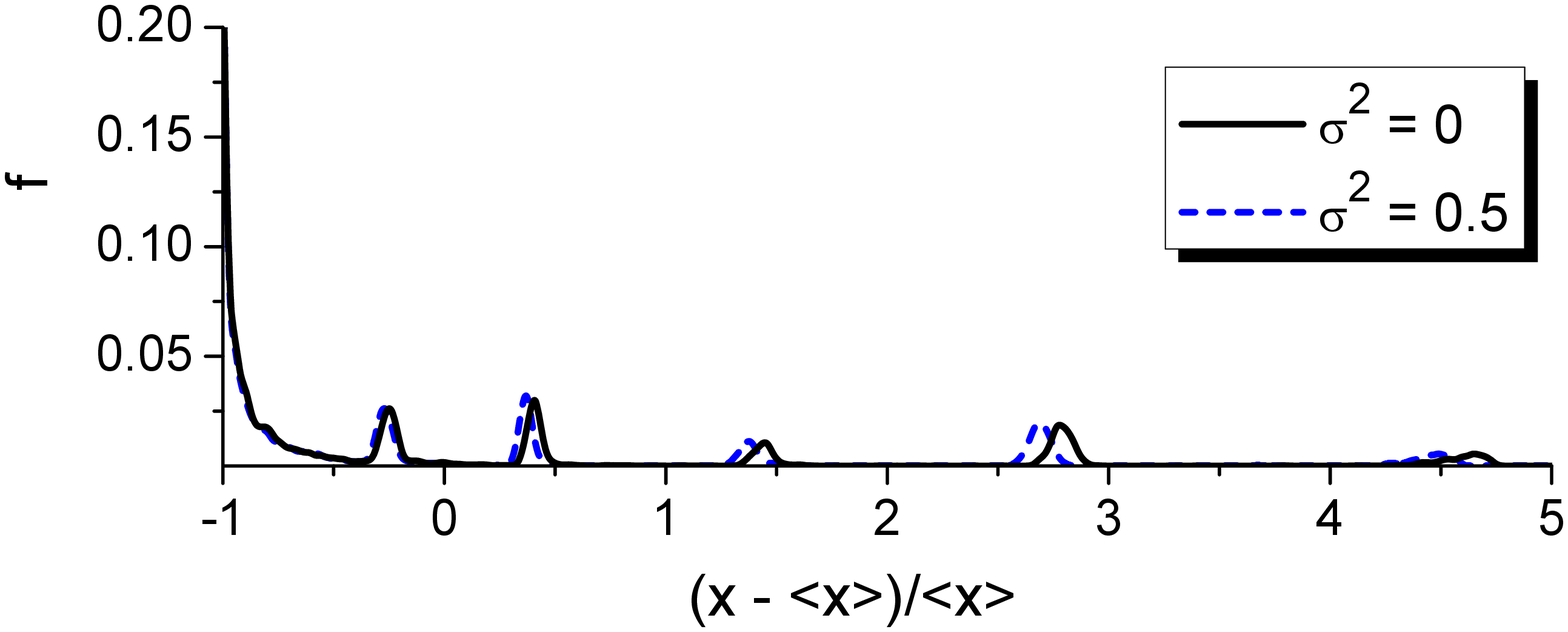}
 \caption{(a) Stationary
distribution of vacancy clusters (at $\sigma^2=0$ and $\sigma^2=0.5$) and
density of vacancy clusters \emph{versus} noise intensity $\sigma^2$. (b)
Structure function for vacancy concentration field at deterministic and
stochastic conditions ($\sigma^2=0.1$). (c) Distribution functions over vacancy
clusters occupancy at deterministic and stochastic conditions Other parameters
are: $T=900K$ and $K=10^{-3}~dpa/sec$. \label{voidnumber}}
\end{figure}

Comparing distributions of vacancy clusters by the size $s$ in deterministic
and stochastic cases it follows that external fluctuations lead to vacancy
migration between clusters. As a result $s$ attains the mean value $\langle
s\rangle$ (see top panels in Fig.\ref{voidnumber}). The noise action does not
influence on vacancy clusters densities $n/N^2$ essentially, the number of
clusters slightly grows with an increase in $\sigma^2$.
Considering behavior of the spherically averaged structure function of the
vacancy concentration field shown in Fig.\ref{voidnumber}b it follows that the
external noise shifts slightly the peak position leading to formation of
clusters distributed with small period. With the noise intensity growth the
area under the curves $S(k)$ increases that well corresponds to elevated values
of the order parameter $\langle(\delta x)^2\rangle$ in the stochastic case.
Comparing data for the period of vacancy clusters $L_0$ related to the same
data for irradiation under reactor conditions one gets $L_0\sim
(4.5\div4.2)\chi L_d$. The lower value of $L_0$ relates to large noise
intensities. In other words vacancy clusters emerging at irradiation under
accelerator conditions decrease down to $6.25\div 6.6\%$ under the same
stochastic conditions. For the averaged linear size of vacancy clusters one has
$\langle d_0\rangle\simeq 1.25 \chi L_d\simeq 7.5~nm$. The distribution
function $f(x)$ of vacancy clusters occupation is shown in
Fig.\ref{voidnumber}c for both deterministic and stochastic cases. In contrast
to the previously studied case here the most of clusters are characterized by
elevated vacancy concentration $x>\langle x\rangle$. It is principally
important that the external fluctuations act in opposite manner comparing to
the stochastic case discussed above. Indeed, here due to formation of mostly
spherical clusters the stochastic contribution decreases vacancy concentration
in clusters shifting main peaks of $f(x)$ toward small $x$. It is interesting
to note that the maximal value of the distribution function in the point $x=0$
is larger by an order comparing to the same value under reactor conditions. It
means that irradiation under accelerator conditions results to fast motion of
vacancies to clusters leading to an emergence of a dense state of material
without vacancies: the diffusion processes are suppressed here and all
vacancies are collected in clusters.

\begin{figure} \centering
\includegraphics[width=0.95\columnwidth]{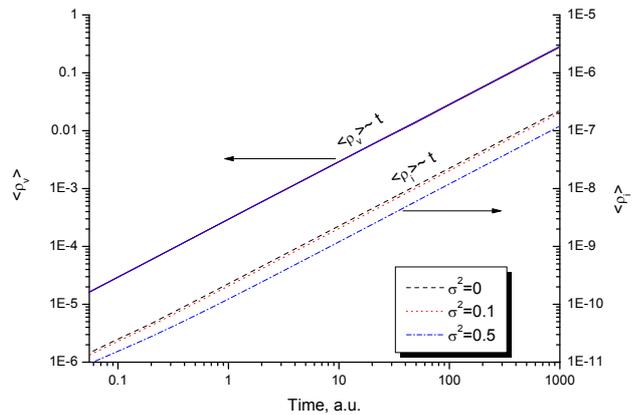}
\caption{Dynamics of sink densities at different values for the noise intensity
 $\sigma^2$ at $T=900K$ and $K=10^{-3}~dpa/sec$.
\label{acc_rirv}}
\end{figure}

Studying dynamics of sink densities (see protocol shown in Fig.\ref{acc_rirv})
one finds here that the linear growth law is observed, i.e.,
$\langle\rho_v(t)\rangle, \langle\rho_i(t)\rangle\propto t$. Comparing behavior
of the sink densities at irradiation under reactor conditions with data shown
in Fig.\ref{acc_rirv} it follows that at large defect production rate $K$ the
universal growth law is realized even at initial stages of the system
evolution. There is no delayed dynamics caused by diffusion processes. The
external fluctuations decrease values for sink densities as in previous case
without delaying dynamics. At large $K$ one gets sink densities by three orders
larger than under reactor conditions.

\section{Conclusions}
In this paper we have studied self-organization processes of vacancy clusters
in stochastic system of point defects subjected to irradiation under reactor
and accelerator conditions. As a prototype physical model the pure nickel with
initial equilibrium vacancy concentration is used. We have considered a general
case by taking into account dynamics of interstitial and vacancy loops. Within
the framework of analytical and numerical analysis we have shown that free
vacancies can arrange in defect clusters of nano-meter size. Obtained results
are in good correspondence with experimental data (see
Refs.\cite{JES,Kiritani00,SYMK2004,Mazias}) and theoretical predictions
\cite{Walgraef95,Walgraef,mirz1}.

Using mean field approximation for the one-component system we have found
critical values of both irradiation temperature and damage rate limiting
self-organization of vacancies into clusters. From obtained numerical data we
conclude that spatial arrangement of defects under both reactor and accelerator
conditions differs by spatial modulation of vacancy concentration and sink
densities fields. From the other hand values of the corresponding averaged
quantities differ by three orders. The universal linear growth law for sink
densities is observed at early stages of the system evolution due to high speed
of point defects arrangement at high damage rates. It is principle important
that in both studied cases the external fluctuations accelerate spatial
organization of defects. However, under reactor conditions the external noise
decreases the averaged point defects concentration, whereas at irradiation in
accelerators such noise increases the corresponding value. In the stochastic
case one has lower sink densities than in deterministic case. The noise delays
dynamics of the sink densities essentially under irradiation in reactor
conditions. We have found that period of vacancy clusters decreases with the
noise intensity growth. Moreover, comparing the corresponding data it was shown
that irradiation at elevated temperatures and high defect production rate
decreases the period of vacancy clusters down to $6.5\%$. From obtained results
for linear sizes of vacancy clusters it follows that the irradiation in
accelerators leads to decrease of the linear size of vacancy clusters comparing
to irradiation in reactors down to $20\%$. We have found that this
characteristic does not depend on the intensity of external fluctuations. Such
difference in linear sizes (period of structures and their diameter) at
irradiation under reactor and accelerator conditions relates to large
difference in defect production rate leading to different speeds of the
corresponding diffusion processes. The same situation is clearly seen from
dynamics of defect sinks. Studying distribution functions of vacancy clusters
occupation we have found that one has fast organization of vacancy clusters
with dense state of material without free vacancies under accelerator
conditions due to suppressed diffusion processes comparing to the case of
irradiation under reactor conditions. The external noise increases the vacancy
concentration in clusters under reactor conditions, whereas it decreases an
occupation of vacancy clusters at irradiation in accelerators.

\end{document}